\documentclass{ws-p8-50x6-00}

\def\APJ{\em ApJ}
\def\APJS{\em ApJS}
\def\AA{\em A\&A}
\def\MNRAS{\em MNRAS} 
\def\PASP{\em Publ. Astr. Soc. Pac.}

\begin{document}

\title{Nucleosynthesis in classical novae}

\author{Margarita Hernanz}

\address{Institut d'Estudis Espacials de Catalunya, IEEC, and 
Instituto de Ciencias del Espacio (CSIC), 
Edifici Nexus, C/Gran Capit\`a, 2-4, E-08034 Barcelona, Spain\\
%E-mail: hernanz@ieec.fcr.es
}

\author{Jordi Jos\'e}

\address{Institut d'Estudis Espacials de Catalunya, IEEC, and 
Departament de F\'{\i}sica i Enginyeria Nuclear (UPC), 
Avda. V\'{\i}ctor Balaguer, s/n, E-08800 Vilanova i la Geltr\'u (Barcelona),
Spain\\
%E-mail: jjose@ieec.fcr.es
}  

\author{Alain Coc}

\address{Centre de Spectrom\'etrie Nucl\'eaire et de Spectrom\'etrie de 
Masse, IN2P3-CNRS, 
Universit\'e de Paris Sud, B\^atiment 104, F-91405 Orsay Cedex, France\\
%E-mail: coc@csnsm.in2p3.fr
}

\maketitle

\abstracts{A general review of the relevance of classical novae for the 
chemical evolution of the Galaxy, as well as for Galactic radioactivity 
is presented. A special emphasis is put on the pioneering work done by Jim 
Truran in this field of research. The impact of recent developments in 
nuclear astrophysics on nova nucleosynthesis, together with the prospects 
for observability of novae radioactivities with the INTEGRAL satellite are 
discussed.}

\section{Introduction}
Classical novae are explosions occurring on the top of white dwarfs accreting 
hydrogen-rich matter in close binary systems, of the cataclysmic variable 
type. The 
transfer of matter results from Roche lobe overflow from the main sequence 
companion star. The accreted hydrogen is compressed up to degenerate 
conditions, leading to a thermonuclear runaway. Explosive hydrogen burning 
occurs via the CNO cycle, out of equilibrium because radioactive nuclei with 
timescales longer than the evolutionary timescale are produced. These nuclei 
are transported by convection to the outer envelope, where they are preserved 
from destruction until they decay. This late decay is the final responsible 
of the envelope expansion and increase in luminosity characteristic of nova 
outbursts. 

It is important to stress that Jim Truran, with Sumner Starrfield and 
collaborators, was a pioneer in the study of thermonuclear explosions. 
Starrfield, Truran, Sparks and Kutter wrote a series of papers, starting 
in the early 70's, which presented the first hydrodynamic computations of 
classical 
novae~\cite{Sta72,Sta74a,Sta74b,Spa76,Sta78}. Models had 
the accreted envelope ``in place'', i.e, the accretion 
phase was not followed. One of the main conclusions was that the envelope 
should be enhanced in CNO nuclei in order to power the nova outburst. This is 
still a valid fact, and it poses severe problems to the current theory, 
because there isn't a clear understanding of the mixing mechanism responsible 
for that enhancement. In 1978, another group also 
developed a hydro code to study nova explosions~\cite{Pri78}, with similar 
limitations 
concerning the accretion phase. It is also worth mentioning that Jean Audouze 
also contributed with Jim Truran to the progress in the nova field, since he 
studied the properties of the hot CNO burning~\cite{Aud73,Laz79}, 
which (although not ``so hot'') is crucial for the explosion.

Classical nova explosions are very common: $\sim$35 per year in our Galaxy, 
from which only 3 to 5 are discovered. They release large amounts of energy, 
$\sim 10^{45}$ erg, but not large enough to have an impact on the dynamics 
of the interstellar medium, like supernovae. The small mass ejected (in the 
range 
$10^{-5}-10^{-4}$ M$_\odot$) implies that novae scarcely contribute to the 
chemical evolution of the Galaxy. If we adopt an ejected mass of 
$2\times10^{-5}$M$_\odot$, a Galactic nova rate of 35 yr$^{-1}$ and an age 
of the Galaxy of $10^{10}$ yr, novae can account only for $\sim 1/3000$ of the 
gas and dust in the Galactic disk; this number is somehow a lower limit, 
since the ejected mass deduced from observations of some novae is larger 
than the typical theoretical value adopted here. 
However, novae are the main producers of 
some particular isotopes ($^{13}$C, $^{15}$N and $^{17}$O), they can help 
to explain the $^{7}$Li versus metallicity relation in the Galaxy 
and they contribute to the radioactivity of the Galaxy, through the ejection 
of medium and long-lived radioactive isotopes (such as $^{7}$Be, $^{22}$Na and 
$^{26}$Al). In addition, novae often form dust, which manifests through their 
IR emission; therefore they could explain the anomalous isotopic 
signatures measured in some presolar grains~\cite{Ama01,Jos01} (see also 
the excellent recent review~\cite{Geh98}). 

\section{Novae and the chemical evolution of the Galaxy}
Novae can be classified in two types, according to their observed abundances. 
The often called ``standard'' novae are objects where enrichments in CNO 
nuclei have been reported, whereas the so-called ``neon'' novae show also 
enrichments in neon. This class of novae was first studied hydrodynamically 
by Starrfield et al.~\cite{Sta86}, including the follow-up of the accretion 
phase. The two nova types are interpreted as the result of 
explosions on accreting carbon-oxygen (CO) and oxygen-neon (ONe) white 
dwarfs\footnote{Until recently, massive white dwarfs were 
called ONeMg white dwarfs, because they were thought to be made of these 
elements. But now it is known that these stars are almost devoid of magnesium 
and, consequently, should be called ONe white dwarfs}, 
respectively, because the chemical composition of the ejecta 
reflects that of the underlying white dwarf. The reason is that some mixing 
between the accreted envelope (presumed to have close to solar composition) 
and the underlying core is required, in order both to power the explosion 
and to explain the observed enrichments in CNO and/or Ne. 
The question of the frequency of occurrence of ONe white dwarfs in classical 
novae was addressed by Truran and Livio~\cite{Tru86,Liv94}, who concluded 
that these white dwarfs can account for $\sim 1/3$ of all observed outbursts.

The mechanism 
responsible of the mixing has not been determined yet (shear mixing, 
diffusion, convection). Recent 
two and three-dimensional calculations~\cite{Gla97,Ker98,Ker99} address the 
convective 
mixing during the hydrodynamic phase following degenerate ignition of 
hydrogen. There isn't yet a clear way to mix and probably the mixing, or some 
part of it, should proceed in the previous hydrostatic accretion phase. 
Diffusion during a number of succesive flashes has been also 
simulated~\cite{Pri95,Kov97}. In this case, moderate 
enhancement can be obtained 
for some combinations of initial conditions (mass, luminosity, accretion 
rate); the remaining problem of this scenario is the handling of mass-loss 
during succesive flashes.

Until some clear mechanism is identified and understood, we~\cite{Jos98} 
adopt a parametrization of the mixing with the underlying core (ranging 
from 25 to 75 \%, 
in order to explain the range of observed metallicities in novae ejecta; 
other authors~\cite{Pol95,Sta98} adopt only 50\%).
In table 1 we show the main nucleosynthetic products of a handful of nova 
models, representative of both classes of novae (see Jos\'e and 
Hernanz~\cite{Jos98} for a larger sample). 
CO white dwarfs have masses up to $\sim 1.15$ M$_\odot$, whereas ONe have 
larger masses (the exact limit between both types of degenerate cores, 
related to the previous evolution during the AGB phase, is still an open 
issue). The accretion rate adopted is $2\times 10^{-10}$ M$_\odot$.yr$^{-1}$ 
and the initial luminosity $10^{-2}$ L$_\odot$ (initial mixing 50 \%). 
We show the ejected masses of the 
most overproduced isotopes ($^{13}$C, $^{15}$N and $^{17}$O), together with 
ejected masses of medium and long-lived radioactive nuclei. 
In figure 1 we 
show the overproduction factors, with respect to solar, of all the isotopes 
contained in the ejecta of a CO and an ONe novae, of the same mass, in order 
to illustrate the impact of the chemical composition of the white dwarf. 
$^{13}$C, $^{15}$N and $^{17}$O are overproduced in both nova types by 
similar factors. ONe novae also eject Ne, required to fit the observations. 
The radioactive isotopes $^{22}$Na and $^{26}$Al are almost only produced in 
ONe novae, because of the presence of seed nuclei in the accreted envelope 
mixed with the core; this presence is necessary to power the NeNa-MgAl 
reaction rate cycles responsible of the synthesis of $^{22}$Na and $^{26}$Al
synthesis. On the other hand, $^{7}$Li (coming from the radioactive $^{7}$Be) 
is produced in larger amounts in CO novae, because of the shorter duration 
of the accretion phase (with T smaller than $\sim 10^8$ K), which is the 
critical one for $^{3}$He survival, with $^{3}$He being essential for $^{7}$Be 
synthesis (see Hernanz et al.~\cite{Her96} for details). The production of 
$^{7}$Li 
in novae was already predicted by the computations of Starrfield et 
al.~\cite{Sta78b}  with their hydro code more than 20 years ago, but at the 
epoch they didn't follow the 
accretion phase, since they had the accreted envelope in place; therefore, 
a crucial phase for $^{7}$Li synthesis was missed.
The relevance of novae for the Galactic content of $^{7}$Li is still an open 
issue. A rough estimate of their average contribution gives a maximum of the 
20\%, 
but the important fact is at which stage of the Galactic evolution they 
contribute. In a recent paper~\cite{Rom99}, it has been shown 
that $^{7}$Li contribution from novae is required in order to reproduce the 
shape of $^{7}$Li versus metallicity relationship, which grows at low 
metallicities. 

Comparison between theoretical abundances in the ejecta and observed 
abundances in particular novae gives a quite good agreement~\cite{Jos98} 
both for CO and ONe novae~\cite{Sta98}, as 
well as a global comparison of average properties in CO novae~\cite{Kov97}. 

\section{Novae and Galactic radioactivity}
The content of CO and ONe novae ejecta in radioactive isotopes is shown in table 1 
for some significative models. As mentioned before, CO novae are more successful in 
producing $^{7}$Be, whereas ONe novae produce larger quantities of $^{22}$Na and 
$^{26}$Al than CO novae. This fact has important consequences for the $\gamma$-ray 
signatures expected for both types of novae, as shown in detail in our recent 
papers~\cite{Gom98,Her99}: individual CO novae will preferentially 
emit photons of 478 keV (with $\tau$=77 days), whereas ONe will emit 1275 keV photons 
($\tau$=3.75 yr). Concerning the cumulative emission, ONe novae can contribute to the
Galactic content of $^{26}$Al ($\tau=10^6$ yr), although their contribution is not the 
most important one: according to the observations of the Galactic 1809 keV emission, 
made with the CGRO/COMPTEL instrument, this emission seems to be preferentially associated 
with a massive star population. 

In the recent years, we have studied in detail the impact that uncertainties in nuclear reactions 
have on nova yields, specially for the radioactive ones ~\cite{Jos99,Coc00}, in order to make 
predictions of detectability of classical novae by the future ESA satellite INTEGRAL, to be 
launched in April 2002. Novae (ONe) up to distances of $\sim 2$kpc could be detected through the 
$^{22}$Na line at 1275 keV, whereas CO novae should be at shorter distances to be detected 
through the $^{7}$Be line at 478 keV~\cite{Gom98}. The most powerful emission of $\gamma$-rays ç
from novae is the line emission at 511 keV and the continuum at energies between 20 and 511 keV, 
related to e$^-$-e$^+$ annihilation, with the positrons coming from the very short-lived 
isotopes $^{13}$N and $^{18}$F. This emission could be detected up to 3 kpc with INTEGRAL, but 
it has very short duration and appears very early, even before optical detection. Detection of novae 
with INTEGRAL would provide a direct proof of the thermonuclear runaway model.
\vspace{-0.2cm}

% Table of nucleosynthesis in novae ejecta
\begin{table}
\begin{center}
\footnotesize
\caption{Nucleosynthesis in CO and ONe novae}
\renewcommand{\arraystretch}{1.1}
\begin{tabular}{|c|c|c|c|c|c|c|c|}
\hline
Nova        & M$_{\rm wd}(\rm M_\odot)$ & $^{13}$C (M$_\odot$)           & 
              $^{15}$N (M$_\odot$)      & $^{17}$O (M$_\odot$)           & 
              $^{7}$Be (M$_\odot$)      & $^{22}$Na (M$_\odot$)          &
              $^{26}$Al (M$_\odot$)\\
\hline
CO          & 0.8                       & 7.7x10$^{-6}$                  &
              9.9x10$^{-8}$             & 2.6x10$^{-7}$                  &
              6.0x10$^{-11}$            & 7.4x10$^{-11}$                 &
              1.7x10$^{-10}$\\
CO          & 1.15                      & 1.3x10$^{-6}$                  &
              5.4x10$^{-7}$             & 2.7x10$^{-7}$                  &
              1.1x10$^{-10}$            & 1.1x10$^{-11}$                 &
              6.1x10$^{-10}$\\
ONe         & 1.15                      & 8.0x10$^{-7}$                  &
              6.8x10$^{-7}$             & 7.9x10$^{-7}$                  &
              1.6x10$^{-11}$            & 6.4x10$^{-9}$                  &
              2.1x10$^{-8}$\\
ONe         & 1.25                      & 6.0x10$^{-7}$                  &
              7.7x10$^{-7}$             & 6.9x10$^{-7}$                  &
              1.2x10$^{-11}$            & 5.9x10$^{-9}$                  &
              1.1x10$^{-8}$\\
\hline
\end{tabular}
\end{center}
\end{table}

\begin{figure} % fig 1
\setlength{\unitlength}{1cm}
\begin{picture}(15,10)
%\put(1,0){\makebox(8,10){\epsfxsize=9cm \epsfbox{fig1.eps}}}
%\put(7.5,0){\makebox(8,10){\epsfxsize=9cm \epsfbox{fig2.eps}}}
\put(0,0){\makebox(8,10){\epsfxsize=9cm \epsfbox{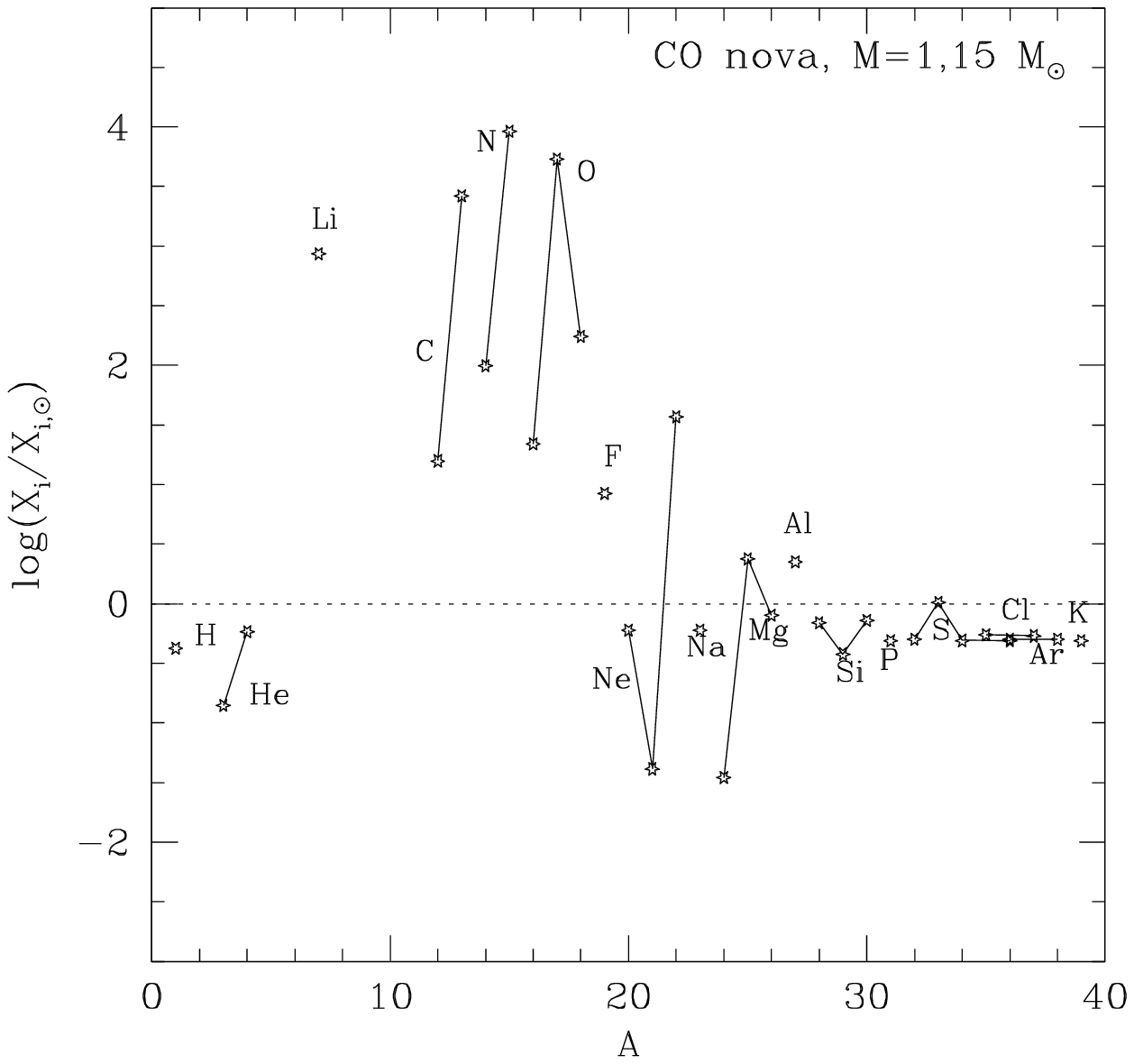}}}
\put(6,0){\makebox(8,10){\epsfxsize=9cm \epsfbox{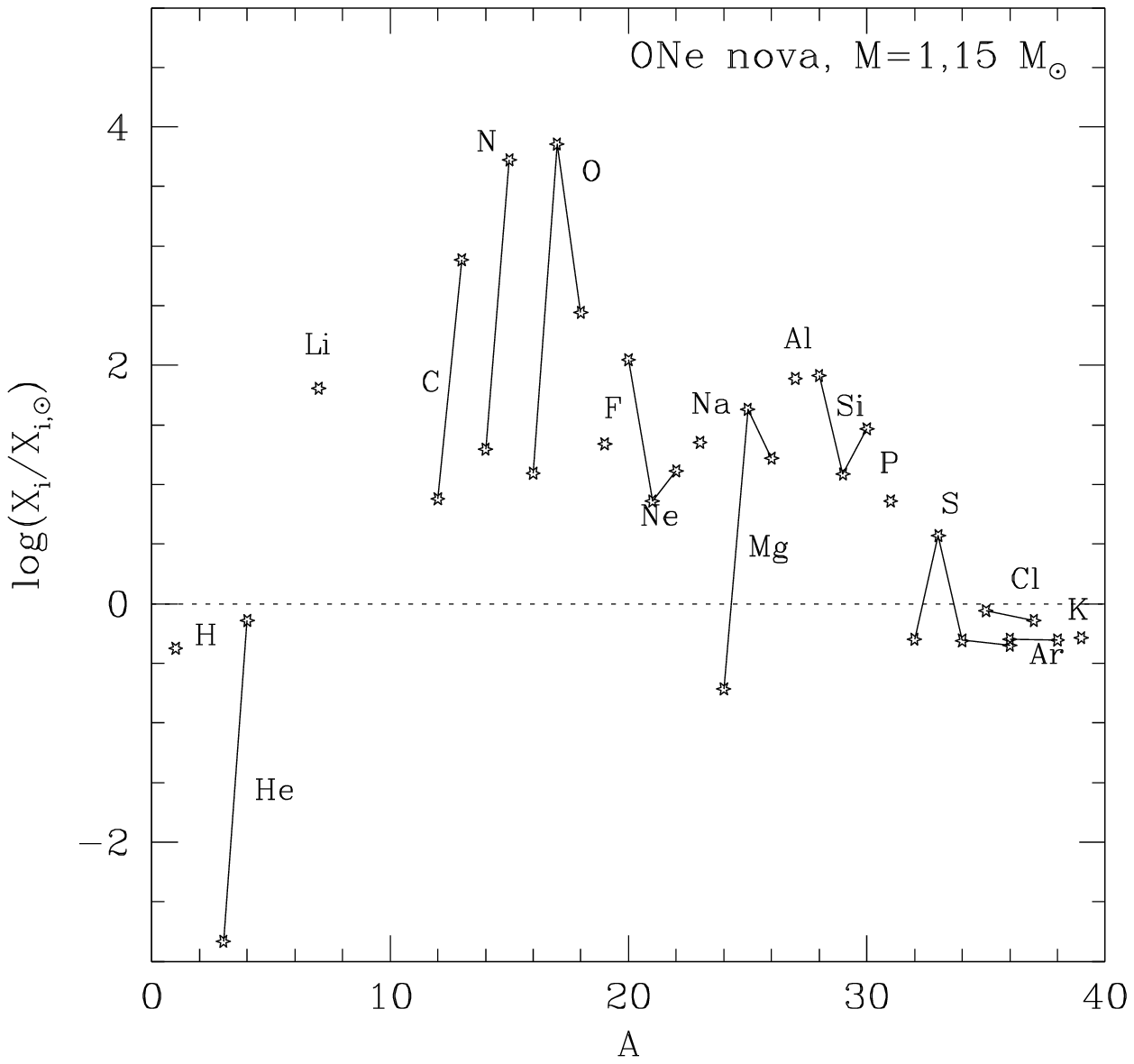}}}
\end{picture}
\vspace*{-2cm}
\caption{Overproduction factors relative to solar abundances, versus 
mass number for: (left) CO nova with M=1.15M$_\odot$, (right) ONe nova 
with M=1.15M$_\odot$}
\end{figure}

\section*{Acknowledgments}
\vspace{-0.2cm}
Research partially supported by the CICYT
(ESP98-1348, PB98-1183-C03-02 and PB98-1183-C03-03) and 
by the AIHF1999-0140.

\end{document}